\documentclass[aps,prl,reprint,superscriptaddress,nofootinbib,floatfix]{revtex4-2}
\usepackage{amsmath,amssymb,bm,graphicx,booktabs,xcolor}
\usepackage[colorlinks=true,linkcolor=blue!50!black,citecolor=blue!50!black,urlcolor=blue!50!black]{hyperref}
\newcommand{\SM}[1]{[SM, Sec.~\ref{SM-#1}]}
\newcommand{\bz}{\bm z}
\newcommand{\bv}{\bm v}
\newcommand{\bh}{\bm h}
\newcommand{\bq}{\bm q}
\newcommand{\bx}{\bm x}
\newcommand{\bo}{\bm o}
\newcommand{\be}{\bm e}
\newcommand{\bc}{\bm c}
\newcommand{\ii}{\mathrm{i}}
\definecolor{alexblue}{RGB}{0,0,160}
\definecolor{tbdorange}{RGB}{230,120,0}

\newcommand{\bk}{\bm k}
\newcommand{\bmm}{\bm u}
\renewcommand{\Re}{\mathrm{Re}\,}
\renewcommand{\Im}{\mathrm{Im}\,}
\newcommand{\Meq}[1]{Eq.~(\ref{#1}) of the main text}
\newcommand{\Mfig}[1]{Fig.~\ref{#1} of the main text}

\begin{document}

\title{Everything in its right place: attention is a radio head}

\author{Llu\'is Arola-Fern\'andez}
\author{Aissam Khadraoui}
\author{Alex Arenas}
\affiliation{Departament d'Enginyeria Inform\`atica i Matem\`atiques, Universitat Rovira i Virgili, 43007 Tarragona, Spain}
\date{\today}

\begin{abstract}
Attention, the core of language models, is a digital calculation over tokens: an exponential selects them,
a division normalizes and a positional code sets the order. We show that attention can instead
emerge from the synchronization of a receiver with its context tokens. Each past token is an oscillator whose state carries its content and whose frequency drifts with its
age. The present token selects them by content, aligning an antenna with its query, and by position, setting the frequency of a receiver with its dial. Solving Kuramoto's problem for this receiver yields an attention layer in closed form with two phases: above a critical coupling set by the context, the receiver locks and its weights normalize; below it, it stays silent and its weights sum to less than one. Trained networks use both phases to select, abstain and count, fail where the theory predicts and keep reading far beyond the training length. Each part of attention finds its place in one device, a radio head: it tunes in to what to hear and when, and stays silent when nothing relevant is on air.
\end{abstract}

\maketitle

In a crowded room we can follow one voice and let the others fade~\cite{cherry}. The brain seems to do it by locking:
rhythms of the auditory cortex follow the speech of the attended speaker and not that of the others~\cite{ding,zion},
and areas that exchange information oscillate in step~\cite{singer,fries}. Machines, instead, attend by calculation. In a
language model~\cite{vaswani}, the embedding of each token $j$ is projected onto a query $\bq_j$, a key $\bm k_j$ and a
value $\bv_j$, vectors of the dimension $d$ of a head, and token $i$ reads the tokens before it with softmax attention,
\begin{equation}
\bo_i=\sum_{j\le i}\frac{e^{\bq_i\cdot\bm k_j/\sqrt d}}{\sum_{l\le i}e^{\bq_i\cdot\bm k_l/\sqrt d}}\,\bv_j ,
\label{eq:softmax}
\end{equation}
an exponential of the overlaps between its query and their keys, divided by its sum. The formula knows no order, so
position is added on top, today mostly by rotating queries and keys~\cite{rope}. The recipe works remarkably well, but
its ingredients were chosen rather than derived, and each brings its patches: gates against attention
sinks~\cite{xiao,qiu} and diluted weights~\cite{velickovic}, rescaled codes beyond the trained length~\cite{press,yarn}.
What attention is, beyond the recipe, remains open.

Physics has answered this kind of question before, by turning a computation into what a physical system does on
its own. Memory became the minima of a spin
energy~\cite{hopfield}, learning the equilibrium of a thermal machine~\cite{ackley}, generation the reversal of a
diffusion~\cite{sohl}, and training a process that follows power laws~\cite{kaplan,hoffmann} and crosses phase transitions~\cite{ziyin,rubin,cui,lacasa,intuition}. Attention has come
closest through memory: the modern Hopfield network updates its state exactly as attention does~\cite{krotov,ramsauer},
and so does a gated associative memory at its fixed point~\cite{vivet}. These energies, however, contain the
exponential and the normalization by design. They tell what is read, not when it was written, nor whether anything
should be read at all. Those are questions of time, and time is what oscillators keep.

\begin{figure}[!b]
\centering
\includegraphics[width=\columnwidth]{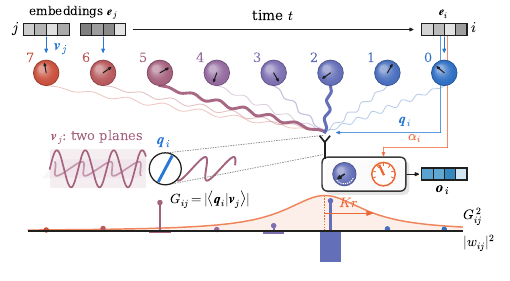}
\caption{The radio head. Each token is a station, numbered and coloured by its age, that carries its value at a frequency falling with that age.
The present token sends its query to the antenna, a polarizer that passes each station with the gain $G_{ij}$ (wave
thickness), and its dial to the receiver, which hears the ages inside its passband (orange, of half width $Kr$). The weight $|w_{ij}|^2$ (bars) combines the squared antenna gain with the resonance response.}
\label{fig:scheme}
\end{figure}

Synchronization is the physics of such questions~\cite{kuramoto,strogatz,arenas}. In a population of coupled oscillators, each oscillator's frequency decides whether it joins the collective rhythm, a self-consistency decides how much it contributes, and below a critical coupling nothing locks. Selection, normalization and silence are thus one
state, which electronic, spintronic and optical oscillators can reach by themselves~\cite{romera,todri}. This makes synchronization a tempting route to attention, and several works have taken it, wrapping oscillatory dynamics around an attention that is still computed~\cite{akorn,nunley,krause,geshkovski} or turning a pairwise locking
condition into weights~\cite{hays}. Pasqualetti and Guo (PG) go furthest, with oscillators that relax on a sphere, from a coupling computed digitally, to an equilibrium that reproduces attention~\cite{pg}. Without natural frequencies, however, oscillators cannot carry position, which then comes from fixed anchors, and an equilibrium normalized by construction can never fall silent.

Here attention is the stationary state of a receiver, an oscillator that listens to the past (Fig.~\ref{fig:scheme}). The past tokens are oscillators too, and their frequency drifts with their age, so that position becomes a frequency. The present token points an antenna at their content with its query, sets the frequency of the receiver with its dial, and the amplitudes of the state it settles into are the attention weights. We find the state, asymptotically exact in wide heads, in closed form with Kuramoto's original insight \cite{kuramoto}: selection, normalization and position follow from one self-consistency condition, which also sets a synchronization threshold below which the receiver falls silent. In trained networks, wide heads make the theory exact, and in them the layer reaches the accuracy of softmax. Furthermore, every receiver can be read like a physical system: its silent phase lets the layer abstain and count, its resonance with the age of the tokens lets it keep reading far beyond the trained length, and its phase boundaries predict where it fails.

\emph{Model.}---We read a head of width $d=2m$ as $\mathbb C^m$, one complex coordinate per pair of real ones, like a light field in $m$ modes, with $\ii^2=-1$, $\langle\bm a|\bm b\rangle=\bm a^\dagger\bm b$ and
$\langle\bm a,\bm b\rangle={\rm Re}\langle\bm a|\bm b\rangle$. Token $j$ is an oscillator $\bx_j$ that carries its content,
$\bv_j=W_V\be_j/|W_V\be_j|$, the unit value of its input $\be_j$, and tells its age by its frequency. Tokens arrive one per unit of a slow time $t$, token $j$ at $t=j$ with zero frequency, which then falls by $\varepsilon$ with every new token. When token $i$ arrives, token $j$ thus turns at $-\varepsilon(i-j)$, redshifted in proportion to its age (Fig.~\ref{fig:scheme}), and keeps this frequency while token $i$ is read, in a much faster time $\tau$,
\begin{equation}
\bx_j=e^{-\ii\varepsilon(i-j)\tau}\bv_j .
\label{eq:age}
\end{equation}

The present token listens to the context in two ways. Its query, the unit vector $\bq_i=W_Q\be_i/|W_Q\be_i|$, points an antenna, which passes token $j$ with the gain
\begin{equation}
G_{ij}=|\langle\bq_i|\bv_j\rangle|^n,
\label{eq:gain}
\end{equation}
louder the better aligned and sharper the higher the directivity $n$, as light through a polarizer (for $n=1$ and $m=2$, $G_{ij}^2$ is Malus' law~\cite{born}). Behind the
antenna sits the receiver $\bz_i$, a unit vector with a dial, its own frequency $-\alpha_i$, which the token sets from its
input, $\alpha_i=\bm w_\alpha^\top\be_i+b_\alpha$. It follows Lohe's equation~\cite{lohe}, Kuramoto's model on the
sphere, turning at the frequency of its dial and pulled, with strength $K$, towards the field $\bh_i$ of the stations it hears,
\begin{equation}
\frac{d\bz_i}{d\tau}=-\ii\alpha_i\bz_i+K\big(\bh_i-r_i\bz_i\big),\qquad \bh_i=\sum_{j\le i}G_{ij}\bx_j,
\label{eq:lohe}
\end{equation}
and settles before the next token arrives. The order parameter $r_i=\langle\bh_i,\bz_i\rangle$, the part of that field in phase with the receiver, is the signal it has caught, and subtracting $r_i\bz_i$ keeps $|\bz_i|=1$. Seen from the receiver, token $j$ turns at the detuning
\begin{equation}
\Delta_{ij}=\alpha_i-\varepsilon\,(i-j),
\end{equation}
which vanishes for the token of age $\alpha_i/\varepsilon$. That token turns with the receiver and resonates, so the dial sets
the age the receiver listens to, as tuning a radio sets the station. The antenna chooses \emph{what} to hear, the dial
\emph{when}. As we derive next, the stationary state of the locked receiver is, in wide heads, a superposition of the tokens, each turning with its
own carrier, $\bz_i=\sum_jw_{ij}\bx_j$, and a silent receiver also keeps a remainder that turns with its dial. The layer reads the amplitudes as a lock-in amplifier
would~\cite{lockin}, each in the frame of its token, and returns
\begin{equation}
\bo_i=g\sum_{j\le i}w_{ij}\bv_j,\qquad w_{ij}=\frac{KG_{ij}}{Kr_i+\ii\Delta_{ij}},
\label{eq:w}
\end{equation}
where $r_i$ is fixed by the unit norm of the receiver above a critical coupling $K_c$, set by the context, and vanishes below it. Each token takes a share $|w_{ij}|^2$ of the receiver, its attention weight, and its value enters the output multiplied by $w_{ij}$, which also turns it by a phase that grows with the detuning. These are the weights of a physical state, with no exponential,
no added positional code and no key: a token is heard by what it carries. In a layer, each token has its own receiver, and all of them are computed at once, like the
rows of softmax. The query, value and output projections $W_Q$, $W_V$ and
 $W_O$, the dial $\bm w_\alpha$, $b_\alpha$ and, per head, $K$, $\varepsilon$, $g$, $n$ and a small floor $\rho$ of the gains are learned by standard backpropagation [Supplemental Material (SM)~\cite{sm},
 Sec.~\ref{SM-sec:impl}]. Table~\ref{tab:dict} translates between synchronization, attention and the radio.

\begin{table}[t]
\caption{Dictionary between synchronization, attention and the radio. Position, selection and normalization follow from one self-consistency condition.}
\label{tab:dict}
\centering
\footnotesize
\setlength{\tabcolsep}{1pt}
\begin{tabular}{@{}lll@{}}
\hline\hline
synchronization & attention & radio\\
\hline
oscillator $\bx_j$ & token $j$, value (no key) & station\\
frequency $-\varepsilon(i-j)$ & position, age $i-j$ & carrier\\
driven part $\sum_jw_{ij}\bx_j$ & reading token $i$ & receiver\\
coupling $KG_{ij}$ & content match & antenna\\
detuning $\Delta_{ij}=\alpha_i-\varepsilon(i-j)$ & position mismatch & dial\\
order parameter $r_i=\langle\bh_i,\bz_i\rangle$ & relevance found & signal\\
complex amplitude $w_{ij}$ & attention weight $|w_{ij}|^2$ & resonance\\
locking width $Kr_i$ & selection & bandwidth\\
locked, $r_i>0$, $\sum_j|w_{ij}|^2=1$ & normalization & gain control\\
silent, $r_i=0$, $\sum_j|w_{ij}|^2<1$ & abstention, counting & squelch\\
demodulated state & $\bo_i=g\sum_jw_{ij}\bv_j$ & audio\\
\hline\hline
\end{tabular}
\end{table}

\emph{Solution.}---We now derive Eq.~\eqref{eq:w} from Eq.~\eqref{eq:lohe}, with a single approximation. Consider one receiver and drop the index $i$. In the frame of the dial, and with dots for $d/d\tau$, Eq.~\eqref{eq:lohe} becomes
\begin{equation}
\dot{\bz}=K(\bh-r\bz),\qquad \bh=\sum_jG_j\bv_je^{\ii\Delta_j\tau},
\label{eq:frame}
\end{equation}
where each token now turns at its detuning. We write the receiver as one mode per token, turning with
it, plus a remainder, $\bz=\sum_jw_j\bv_je^{\ii\Delta_j\tau}+\bc$. Equation~\eqref{eq:frame} then holds when
\begin{equation}
\dot w_j=KG_j-(Kr+\ii\Delta_j)\,w_j,\qquad \dot{\bc}=-Kr\,\bc .
\label{eq:amp}
\end{equation}
Each amplitude is a damped oscillator driven by its token at its detuning, with a friction $Kr$ common to all modes that is the only nonlinearity, because $r$ depends on the state. At constant $r>0$ the amplitudes settle at those of Eq.~\eqref{eq:w}, whatever the contents, with weights on a Lorentzian of half width $Kr$, the band of the receiver, and the remainder decays. Kuramoto's self-consistency closes the loop~\cite{kuramoto,strogatz}: the state
must give back its own $r$ and keep unit norm. Inserting the modes in $r=\langle\bh,\bz\rangle$ gives
\begin{equation}
r=\sum_jG_j\,{\rm Re}\,w_j+{\rm Re}\sum_{j\neq l}G_jw_l\langle\bv_j|\bv_l\rangle e^{\ii(\Delta_l-\Delta_j)\tau},
\label{eq:beats}
\end{equation}
plus terms in $\bc$. The first sum, which pairs each token with its own mode, is constant and equals $r\sum_ja_j$, with $a_j=|w_j|^2$. The second holds the beats between tokens, at the differences of their detunings and with amplitudes set by the overlaps, and the norm $|\bz|^2$ splits in the same way. Averaging the beats out is the only approximation, exact for orthogonal contents. Beats between neighbouring tokens pass the band with a relative amplitude $Kr/\varepsilon$ and act through squared overlaps $\sim1/m$, so the error is of order $Kr/m\varepsilon$.

\begin{figure}[t]
\centering
\includegraphics[width=\columnwidth]{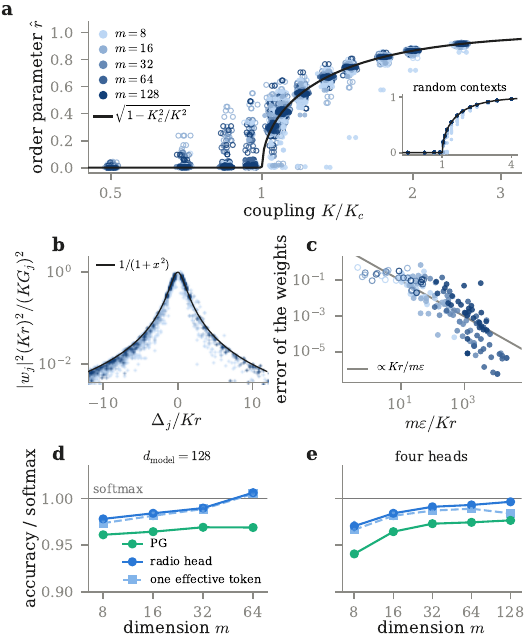}
\caption{The receiver and the layer against $m=d/2$ (colour). (a--c)~Eq.~\eqref{eq:lohe} integrated on
contexts of trained language models; inset of (a), random contexts. (a)~$\hat r$ against the coupling
for resolved ($\varepsilon\ge K_c$, full) and unresolved receivers (open); black, the root of Eq.~\eqref{eq:norm}.
(b)~Rescaled weights of the locked, resolved receivers against $\Delta/Kr$, on the Lorentzian. (c)~Relative error of the
weights against $m\varepsilon/Kr$, one point per receiver at a random locked coupling; grey, fit $\propto Kr/m\varepsilon$.
(d,e)~Next-token accuracy relative to softmax with rotary position for PG, the radio head and its one-token
approximation, with $d_{\rm model}=128$ split into one to eight heads (d) and four heads of growing $d_{\rm model}$ (e). Mean and individual values of two seeds, a single one at $m=128$ in (e) [SM].}
\label{fig:theory}
\end{figure}

\begin{figure*}[t]
\centering
\includegraphics[width=\textwidth]{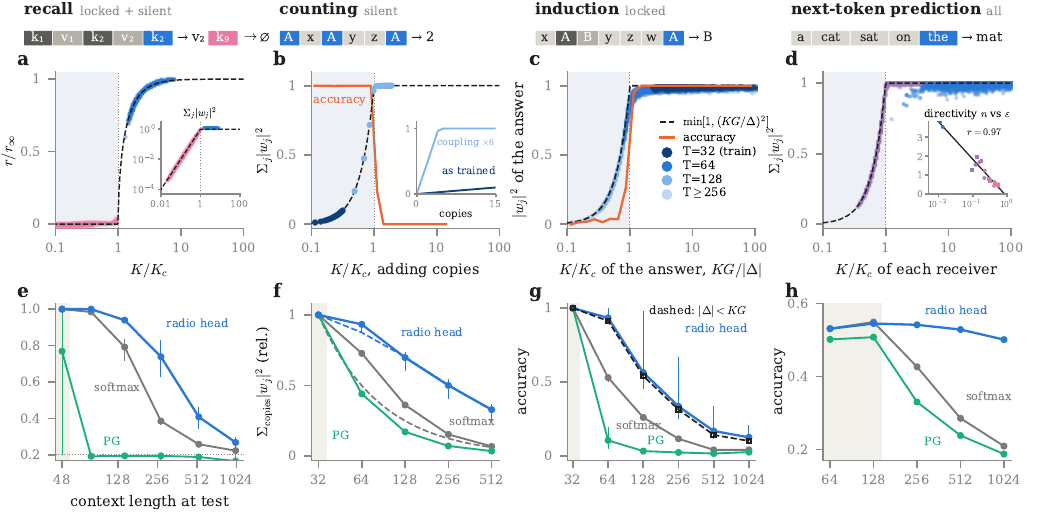}
\caption{The receiver in trained networks (radio head; synthetic tasks: two layers, one head, $d=64$). (a--d)~The
quantity that decides each task against $K/K_c$; dashed, $\min[(K/K_c)^2,1]$ unless stated; shaded, silent side.
(a)~Recall: $r/r_\infty$, with $r_\infty=(\sum_jG_j^2)^{1/2}$ its strong-coupling limit, for listed (blue) and unlisted (pink) keys; dashed, Eq.~\eqref{eq:sqrt}, with $\bar G\simeq r_\infty$ since a single key dominates; inset, total
weight. (b)~Counting: total weight as 1 to 15 copies are added, as trained (dark) and with the coupling raised and the gain compensated (light).
(c)~Induction: weight on the answer against its locking strength $KG/|\Delta|$, from 32 (trained) to 1024 tokens. (d)~Language:
total weight of every receiver, from slow (blue) to fast (pink) heads; inset, directivity $n$ against ageing rate $\varepsilon$, two
seeds. Orange in (b,c), accuracy. (e--h)~Beyond the trained length (shaded), for softmax with rotary position, PG and the radio
head: accuracy in recall (e), weight on the copies relative to the trained length in counting (f; dashed, softmax $32/T$ at context length $T$, and the radio-head prediction), and accuracy in induction (g; dashed, fraction of answers inside the locking range) and language (h). Five seeds (e--g), two (h), mean and full range [SM].}
\label{fig:physics}
\end{figure*}

The two conditions then read $r=r\sum_ja_j$ and $\sum_ja_j+|\bc|^2=1$. Either the receiver locks, $r>0$, the remainder decays and the weights sum to one, a normalization in power rather than in amplitude,
\begin{equation}
\sum_j\frac{K^2G_j^2}{(Kr)^2+\Delta_j^2}=1,
\label{eq:norm}
\end{equation}
which, since its left side falls with $r$, has a root if and only if the coupling exceeds a threshold set by the context,
\begin{equation}
K_c=\Big(\sum_j\frac{G_j^2}{\Delta_j^2}\Big)^{-1/2}.
\label{eq:kc}
\end{equation}
Or the receiver stays silent, $r=0$, and its amplitudes orbit around $w_j=KG_j/\ii\Delta_j$, with weights that sum to
\begin{equation}
\sum_ja_j=\frac{K^2}{K_c^2}=K^2\sum_j\frac{G_j^2}{\Delta_j^2}<1 .
\label{eq:silent}
\end{equation}
The remainder no longer decays and carries the missing norm, $|\bc|^2=1-K^2/K_c^2$, unheard by a layer that reads each token in its own frame. The silent state is a centre exactly when $K<K_c$, so the same threshold separates the two
phases~\SM{sec:stability}. For a single token the locked state is Adler's law~\cite{adler}, $(Kr)^2=K^2G^2-\Delta^2$, inside the locking range $|\Delta|<KG$.
For many, replacing the context by one effective token with the same moments $\sum_jG_j^2/\Delta_j^2$ and $\sum_jG_j^2/\Delta_j^4$, of gain $\bar G=\big(\sum_jG_j^2/\Delta_j^2\big)\big/\big(\sum_jG_j^2/\Delta_j^4\big)^{1/2}$, gives
\begin{equation}
r=\bar G\sqrt{1-K_c^2/K^2},
\label{eq:sqrt}
\end{equation}
asymptotically exact near the onset, where $r$ rises with Kuramoto's square root~\SM{sec:onset}. The layer never integrates the
dynamics. It computes $Kr$ from Eq.~\eqref{eq:norm} by bisection, or from the effective token, and is
trained by backpropagation through the closed form. At $K_c$ the derivatives of $r$ diverge, so every
token carries a weak phase noise, a linewidth $D=10^{-3}K$ that rounds the transition and keeps the gradients
finite~\SM{sec:linewidth}.

Figure~\ref{fig:theory}(a--c) tests the closed form against the dynamics. We train a small transformer on TinyStories~\cite{eldan} with the radio head in place of softmax, take the gains, detunings and contents of its receivers, and integrate Eq.~\eqref{eq:lohe} itself across the threshold~\SM{sec:check}. Since resolution in frequency needs $\varepsilon>Kr$, and $Kr$ stays of order $K_c$ over the sweep, we call a receiver resolved when $\varepsilon\ge K_c$, fixed before any comparison, and show both kinds. To compare contexts we use $\hat r^2=1-K_c^2\sum_jG_j^2/[(Kr)^2+\Delta_j^2]$, which equals $1-K_c^2/K^2$ at the root. Resolved receivers follow the root of Eq.~\eqref{eq:norm} [Fig.~\ref{fig:theory}(a)], their weights fall on a single Lorentzian [Fig.~\ref{fig:theory}(b)], and the error that remains falls as $Kr/m\varepsilon$, as predicted [Fig.~\ref{fig:theory}(c)]. Slow heads, which hear neighbours as one, still follow the closed form within 10--20\%, and silent receivers orbit around the state the layer reads, on times short compared with $1/D$~\SM{sec:check}.

\emph{Trained networks.}---Wide heads also make the layer accurate. With the same protocol for every attention layer~\SM{sec:impl}, whether heads widen at fixed model size or the model grows, the radio head approaches softmax with rotary position and matches it in the widest heads, while the oscillator layer of PG stays behind. The one-token approximation, exact near the onset, loses ground when heads lock deeply [Fig.~\ref{fig:theory}(d,e)]. Its phase diagram also makes predictions that a normalized layer cannot meet~\SM{sec:props}. A
query that finds nothing should abstain, copies below threshold should add up instead of sharing a budget, a token should
be heard only inside the locking range, and a receiver tuned to an age should keep hearing it at any context length. We test which phase training chooses, and whether its boundaries predict where the network fails, on three small tasks, each needing a different part of the phase diagram, and on language,
which needs all of it (Fig.~\ref{fig:physics}).

\emph{Abstention.}---In recall~\cite{arora} the network sees key--value pairs and is asked for the value of a
key, sometimes absent from the list. It must lock to answer and stay silent to abstain, and trained receivers do both, on the curve of the onset [Fig.~\ref{fig:physics}(a)]. The weight a silent receiver does not give stays in its remainder, so no sink token is needed~\cite{xiao,qiu}. Given more pairs than in training, the radio head degrades slowly, softmax
faster, and PG falls to chance [Fig.~\ref{fig:physics}(e)].

\emph{Counting.}---To count how often the query has appeared before, the network needs silence. Below threshold the
total weight, $(K/K_c)^2=K^2\sum_jG_j^2/\Delta_j^2$, is a sum over tokens, so each copy of the query adds its own term, set by its age. Trained to count up to 15 copies, the network keeps its receivers far below threshold. Raising the
coupling, with the gain lowered to leave the silent layer unchanged, brings the threshold within reach, and the total weight grows linearly until it saturates at one, where the
count is lost [Fig.~\ref{fig:physics}(b)]. Softmax without a sink always sums to one, so it can only count the
fraction of the context taken by the copies, which falls with the length, whereas the weight of the radio head falls only as
the copies age, as the Lorentzian predicts [Fig.~\ref{fig:physics}(f)].

\emph{The locking range.}---Induction~\cite{olsson}, finding the previous occurrence of the current token and copying what
followed it, needs locking alone. The first layer becomes a clock tuned to the previous token, and the second
finds the answer by its content. Its weight on the answer is that of a single station, full inside Adler's
locking range and falling as $(KG/\Delta)^2$ outside it, and the accuracy collapses at that edge [Fig.~\ref{fig:physics}(c)].
On longer contexts the answers drift out of the locking range, and the fraction that stays inside predicts the
accuracy at every length [Fig.~\ref{fig:physics}(g)]. The network fails where the theory says it must, and this holds across seeds~\SM{sec:results}.

\emph{Language.}---Next-token prediction needs all of it. Most receivers lock and some stay silent
[Fig.~\ref{fig:physics}(d)]. Training spreads heads that start with the same ageing rate into a ladder of time
scales, the same in every seed, and heads that listen further sharpen their antennas. A slow head holds about $Kr/\varepsilon$ tokens in its band, each unrelated one
with squared gain $m^{-n}$, and singles out the one it seeks only if $n\gtrsim\ln(Kr/\varepsilon)/\ln m$~\SM{sec:capacity}.
Consistently, the learned directivity grows linearly with $\ln(1/\varepsilon)$ [inset of Fig.~\ref{fig:physics}(d)]. The slowest
head locks deepest and reads by content alone, the fast ones by position, closer to threshold. Since
resonance depends on the age of a token and not on the length of the context, the radio head nearly matches softmax at the
trained length and keeps its accuracy far beyond it, where softmax with rotary position and PG lose half of it or more
[Fig.~\ref{fig:physics}(h)]. Linear biases~\cite{press} impose such a decay by hand. Here it comes from the resonance: the Lorentzian tail falls as the inverse square of the age, so the sum $\sum_jG_{ij}^2/\Delta_{ij}^2$ that sets each threshold converges and old tokens add a bounded background~\SM{sec:results}.

\emph{Discussion.}---The radio head is a mean-field theory of attention. The closed form neglects only the beats between
tokens, of order $1/m$, so the head dimension controls the accuracy of the theory, much as population size does in
Kuramoto's model. A single phase locks to one token at a time. The dimension is what lets the
receiver hear many at once~\SM{sec:kuramoto}. In narrow or slow heads it becomes the mean field of the dynamics rather than its exact state. Each factor
of Eq.~\eqref{eq:softmax} has a physical counterpart: the exponential of the overlap becomes a polarizer, the positional code a
Lorentzian resonance and the normalization the unit norm of a locked state. Normalization, abstention and counting become three readings of one transition. Because the state is
physical, a trained network can be read before it is tested. Whether the same holds in large models, trained on long and varied text, is
the question these small networks leave open.

All of this ran on digital machines, where the radio head has the same quadratic cost as softmax. Physical oscillators already lock, detune and drift as Eq.~\eqref{eq:lohe}
requires~\cite{romera,todri}, and a receiver relaxing among them would select, normalize and order one read of attention in
a few of its oscillations. The model also separates two ways of attending that neural rhythms may combine: by content and by
time~\cite{fries}. Read this way, attention is the state of a listener in a crowded room, tuned to some voices by what
they say and when they spoke, and silent when nothing it hears is worth locking to.

\begin{acknowledgments}
We thank David Soriano-Pa\~nos and Arnau Vivet for very helpful discussions and a careful reading of the manuscript.
This work has been supported by the Spanish Ministerio de Ciencia, Innovaci\'on y Universidades, PID2024-158120NB-C21.
A.A. also acknowledges the ICREA Academia program of the Generalitat de Catalunya.
We used Claude Opus 5.5 (Anthropic) to write code, prepare figures and work out exploratory and tedious analytical
calculations. The main ideas and results are the authors' own, and they take full responsibility for the content.
\end{acknowledgments}

\emph{Data availability.}---The code and data are available at \url{https://github.com/mystic-blue/radiohead}. 

\clearpage
\onecolumngrid
\setcounter{equation}{0}
\setcounter{table}{0}
\setcounter{figure}{0}
\setcounter{section}{0}
\setcounter{secnumdepth}{2}
\renewcommand{\thesection}{S\arabic{section}}
\renewcommand{\theequation}{S\arabic{equation}}
\renewcommand{\thetable}{S\arabic{table}}
\renewcommand{\thefigure}{S\arabic{figure}}
\makeatletter
\renewcommand{\@biblabel}[1]{[S#1]}
\renewcommand{\citenumfont}[1]{S#1}
\makeatother

\begin{center}
{\large\bfseries Supplemental Material for\\[2pt] ``Everything in its right place: attention is a radio head''}\\[8pt]
Llu\'is Arola-Fern\'andez, Aissam Khadraoui and Alex Arenas
\end{center}
\vspace{6pt}

\noindent This document collects what the letter only states. Section~\ref{SM-sec:km} derives the radio head step by step,
from the equation of the receiver to its two phases. Section~\ref{SM-sec:check} checks the closed form against the integrated
dynamics, and Sec.~\ref{SM-sec:others} compares the layer with softmax and with the layer of Pasqualetti and Guo, and briefly with other layers. Section~\ref{SM-sec:impl}
gives the implementation and the training methods, and Sec.~\ref{SM-sec:results} the checks and controls behind each experiment.
Equations, sections and tables numbered with an S belong to this document.

\section{The radio head}\label{SM-sec:km}

\subsection{Model}\label{SM-sec:model}

A head of width $d=2m$ is read as $\mathbb C^m$, one complex coordinate per plane. The rotation $J$ by a right angle in
every plane is then multiplication by $\ii$. We write $\langle\bm a|\bm b\rangle=\bm a^\dagger\bm b$ for the Hermitian product
and $\langle\bm a,\bm b\rangle=\Re\langle\bm a|\bm b\rangle$ for the real one. In a real network this is only a way of writing
$\mathbb R^{2m}$; in a physical head, such as $m$ modes of light, the field is complex from the start, a token stores its
content in the shape of the field and its age in its phase, and $\langle\bh,\bz\rangle$ is the part of the field in phase
with the receiver, as in a homodyne detector.

The model runs on two times. The slow time $t$ counts tokens: token $j$ arrives at $t=j$. In the much faster time $\tau$
the receiver of token $i$ settles, while the frequencies of the tokens keep their values at $t=i$; its state at
$\tau\to\infty$ (or its average, when it does not settle) is what the layer reads. Every token has its own receiver, which
listens to its causal context $C_i$ ($j\le i$), and all are computed in parallel, as the rows of softmax are.

Token $j$ carries the unit content $\bv_j=W_V\be_j/|W_V\be_j|$, computed from the input $\be_j$ of the block,
which is also its key. It is an oscillator, $\bx_j=e^{\ii\theta_j}\bv_j$, whose frequency depends on its age. It is born at
zero frequency and slows down at the rate $\varepsilon$ of the head, so that at the slow time $t$ it turns at
$\omega_j(t)=-\varepsilon(t-j)$. When token $i$ listens, $t=i$ and every earlier token turns, in the fast time, at the fixed
frequency $-\varepsilon(i-j)$ set by its age. Position is thus a frequency: older tokens sit further down the dial. In
language the context includes token $i$ itself, a station of age 0 that the receiver can hear like any other. In the
synthetic tasks the diagonal is masked, for every layer.

Token $i$ points an antenna along its unit query
$\bq_i=W_Q\be_i/|W_Q\be_i|$ and hears token $j$ with the gain
\begin{equation}
G_{ij}=\big|\langle\bq_i|\bv_j\rangle\big|^{n}.
\label{SM-eq:gain}
\end{equation}
The Hermitian product makes the gain blind to the phase of the token: the antenna reads what a token says, not where its
carrier is. For $n=1$ the gain is the length of the projection of $\bv_j$ on the complex line of $\bq_i$, and for $m=2$ its
square is Malus' law. A learned directivity $n>1$ narrows the pattern. Nothing below depends on the form of the gain.

The receiver $\bz_i\in\mathbb C^m$ is a unit vector, tuned to $-\alpha_i$ by the dial $\alpha_i=\bm w_\alpha^\top\be_i+b_\alpha$. It
follows Lohe's equation~\cite{S-lohe}, \Meq{eq:lohe},
\begin{equation}
\dot{\bz}_i=-\alpha_iJ\bz_i+K\big(\bh_i-\langle\bh_i,\bz_i\rangle\bz_i\big),\qquad \bh_i(\tau)=\sum_{j\in C_i}G_{ij}\,\bx_j(\tau),
\label{SM-eq:lohe}
\end{equation}
over the causal context $C_i$, with $\tau$ in place of $t$ and the frequencies of the tokens frozen at their ages. The
receiver turns at its own frequency and is pulled towards the field of the tokens it hears.

The term $\langle\bh_i,\bz_i\rangle\bz_i$ keeps $|\bz_i|=1$. This constraint is the origin of the normalization, of the
threshold and of the silent phase: a linear receiver would neither normalize nor switch off.

The stationary receiver is a sum of modes, one per token, each turning with the carrier of its token
(Sec.~\ref{SM-sec:amp}). Read directly, $\bz_i$ would mix them with phases that change in time. The layer reads it as a lock-in
amplifier does: it demodulates the mode of every token with the carrier of that token and keeps its complex amplitude
$w_{ij}$. The real part is the part of the mode in phase with its token, the imaginary part the part a quarter of a turn
behind, which is read along $J\bv_j$:
\begin{equation}
\bo_i=g\sum_{j\in C_i}w_{ij}\,\bv_j,\qquad w\bv\equiv(\Re w)\bv+(\Im w)J\bv,
\label{SM-eq:out}
\end{equation}
followed by an output map $W_O$. The amplitudes $w_{ij}$ are derived next.

\subsection{Integration at constant friction}\label{SM-sec:amp}

We first solve the receiver as if the friction were fixed, and impose self-consistency afterwards, as in Kuramoto's
theory. We drop the index $i$ and work in the fast time. In the frame that turns with the dial the term $-\alpha J\bz$ disappears,
and token $j$ turns at its detuning from the dial,
\begin{equation}
\Delta_j=\alpha-\varepsilon(i-j),
\end{equation}
which vanishes for the tokens of age $\alpha/\varepsilon$: a token resonates when its frequency matches that of the dial.
Absorbing a birth phase in $\bv_j$, Eq.~\eqref{SM-eq:lohe} becomes
\begin{equation}
\dot{\bz}=K\big(\bh-r\bz\big),\qquad \bh(\tau)=\sum_jG_j\bv_je^{\ii\Delta_j\tau},\qquad r=\langle\bh,\bz\rangle .
\label{SM-eq:frame}
\end{equation}
The order parameter $r$ is the part of the field in phase with the receiver. It is the only thing that makes the equation
nonlinear. If $r$ were constant, the equation would be linear, $\dot{\bz}+Kr\bz=K\bh$, and an integrating factor gives
$\bz(\tau)=e^{-Kr\tau}\bz(0)+K\int_0^\tau e^{-Kr(\tau-s)}\bh(s)\,ds$. The field is a sum of exponentials, so the integral is done
token by token:
\begin{equation}
\bz(\tau)=\sum_jw_j\bv_je^{\ii\Delta_j\tau}+\bc\,e^{-Kr\tau},\qquad w_j=\frac{KG_j}{Kr+\ii\Delta_j},\qquad
\bc=\bz(0)-\sum_jw_j\bv_j .
\label{SM-eq:w}
\end{equation}
This holds for any number of tokens and any overlaps between their contents. Each token leaves in the receiver a mode
that turns with it. Its amplitude is the response of a damped oscillator driven at the detuning of the token, with the
friction $Kr$ common to all modes. The rest, $\bc$, is the memory of the initial state, and it decays.

Written as $\bz(\tau)=\sum_jw_j(\tau)\bv_je^{\ii\Delta_j\tau}+\bz_\perp$, the same statement is an equation for every amplitude,
\begin{equation}
\dot w_j=KG_j-\big(Kr+\ii\Delta_j\big)w_j,\qquad \dot{\bz}_\perp=-Kr\,\bz_\perp ,
\label{SM-eq:amp}
\end{equation}
a driven, damped oscillator per token, all coupled only through $r$. The amplitudes are well defined when the tokens can be
told apart, in content or in frequency ($\varepsilon>Kr$, the condition that separates resolved from unresolved receivers
in Sec.~\ref{SM-sec:check}).

\subsection{Self-consistency}\label{SM-sec:solution}

The friction is not a constant: $r$ and the norm are produced by the state. As in Kuramoto's theory~\cite{S-kuramoto,S-acebron},
we assume an order parameter, compute the response of every token to it, Eq.~\eqref{SM-eq:w}, and require that the response
gives back the same order parameter and a unit vector:
\begin{equation}
r=\langle\bh,\bz\rangle,\qquad |\bz|^2=1 .
\end{equation}
Inserting Eq.~\eqref{SM-eq:w}, both conditions contain constant terms, one per token, and beats between pairs of tokens. The
beats oscillate at the differences of the detunings, $\Delta_l-\Delta_j$, with amplitudes proportional to the overlaps
$\langle\bv_j|\bv_l\rangle$. We replace them by their time average, zero. This is the method of averaging (the
rotating-wave approximation of optics) and the only approximation of the theory. It is exact for orthogonal contents. For
random contents the overlaps are of order $m^{-1/2}$ and the error of order $|\langle\bv_j|\bv_l\rangle|^2\sim1/m$: the width of
the head plays the role of the number of oscillators in Kuramoto's theory. The beats also need to be faster than the band,
$\varepsilon>Kr$, to average out (Sec.~\ref{SM-sec:check}).

With $a_j\equiv|w_j|^2$ and $G_j\Re w_j=r\,a_j$, the averaged conditions read
\begin{equation}
r=r\sum_ja_j,\qquad \sum_ja_j+|\bc|^2e^{-2Kr\tau}=1 .
\end{equation}
Either $r>0$, the memory decays and the weights sum to one,
\begin{equation}
\sum_j\frac{K^2G_j^2}{(Kr)^2+\Delta_j^2}=1,
\label{SM-eq:norm}
\end{equation}
\Meq{eq:norm}. The left side decreases with $r$, so a root $r>0$ exists, and is unique, if and only if
\begin{equation}
K>K_c=\Big(\sum_j\frac{G_j^2}{\Delta_j^2}\Big)^{-1/2}.
\label{SM-eq:kc}
\end{equation}
Or $r=0$: the receiver is silent, $w_j=KG_j/\ii\Delta_j$ and
\begin{equation}
\sum_ja_j=\frac{K^2}{K_c^2}\qquad(K\le K_c).
\label{SM-eq:silent}
\end{equation}
The memory $\bc$ no longer decays and carries the missing norm, turning at the dial. The total weight, $\min(K^2/K_c^2,1)$, is
continuous at $K_c$. For one token Eq.~\eqref{SM-eq:norm} is Adler's law~\cite{S-adler}, $(Kr)^2=K^2G^2-\Delta^2$.

The weights depend only on ratios. Rescaling $K$, $\varepsilon$ and $\alpha$ together (and the linewidth $D$ of Sec.~\ref{SM-sec:linewidth})
leaves the weights unchanged: $\varepsilon$ only sets the unit of time. A receiver is described by three numbers, the
coupling relative to its threshold, $K/K_c$, the age it listens to, $\alpha/\varepsilon$, and the number of tokens in its band,
$Kr/\varepsilon$.

\subsection{Stability and the silent branch}\label{SM-sec:stability}

The self-consistency gives two branches but does not say which one the dynamics reaches. A single token, where everything
can be done exactly, shows what happens. For one token $r=G\Re w$, and Eq.~\eqref{SM-eq:amp} is exact. With $x=\Delta/KG$, $w=u+\ii\eta$ and
time in units of $1/KG$, it reads
\begin{equation}
\dot u=1+x\eta-u^2,\qquad \dot\eta=-u(x+\eta).
\end{equation}
Dividing the two equations, with $s=x+\eta$, turns them into a linear equation for $u^2$,
$d(u^2)/ds=2u^2/s-2(1-x^2+xs)/s$, whose integral is a conserved quantity,
\begin{equation}
\mathcal I=\frac{u^2-1-x^2-2x\eta}{(x+\eta)^2}.
\end{equation}
For $|x|<1$ the locked state $w=\sqrt{1-x^2}-\ii x$ attracts every start, with the phase lag $\arcsin x$. For $|x|>1$ there is
no locked state, and the orbits are the level curves of $\mathcal I$ around the silent state $w^*=-\ii/x$. Their demodulated
power $|\langle w\rangle|^2$ lies between $1/x^2$, at the centre, and $(|x|-\sqrt{x^2-1})^2$ on the outermost orbit. On that orbit
$|w|=1$ and the remainder $\bc$ vanishes: the receiver is a pure phase that drifts as in Adler's equation [Eq.~\eqref{SM-eq:adler}].
Since $\dot{\bc}=-Kr\,\bc$, the orbit without remainder is invariant, and a weak phase noise in the token carries the receiver
onto it: in a simulation of one token with noise, $|w|$ tends to one and stays there. This drift is slow, over times of order $1/D$, a thousand times the time unit $1/K$ of the layer (Sec.~\ref{SM-sec:linewidth}). The layer reads the orbit reached from a nearly orthogonal start well before, Eq.~\eqref{SM-eq:orbit}, and the linewidth regularizes this reading rather than describing the drift. The weight of one token at the
centre, $\min[1,(KG/\Delta)^2]$, is the dashed curve of Fig.~\ref{fig:physics}(c) of the main text.

With many tokens, the silent state $w_j=KG_j/\ii\Delta_j$ is a fixed point of Eq.~\eqref{SM-eq:amp} with $r=0$. Linearized there,
$\delta\ddot x=-(\Lambda^2-K^2\bm g\bm g^\top)\,\delta x$, with $x_j=\Re w_j$, $\Lambda={\rm diag}(\Delta_j)$ and $\bm g=(G_j)$. By the matrix
determinant lemma, $\det(\Lambda^2-K^2\bm g\bm g^\top)=\det(\Lambda^2)(1-K^2/K_c^2)$, and a rank-one perturbation of a positive diagonal
matrix has at most one negative eigenvalue. So the silent state is a centre exactly when $K<K_c$, and one frequency
vanishes at $K_c$: the receiver slows down near threshold. The stability of the locked state with many tokens is checked numerically (Sec.~\ref{SM-sec:check}).

Below threshold the layer still needs one value to read, and the orbits of many tokens can be followed in perturbation
theory. A random start is nearly orthogonal to the tokens for large $m$, so we start from $w_j(0)=0$ and expand
Eq.~\eqref{SM-eq:amp} in $K$. At first order the friction plays no role,
$w_j^{(1)}=(KG_j/\ii\Delta_j)(1-e^{-\ii\Delta_j\tau})$, whose average is the centre, and its in-phase part gives
$r^{(1)}=\sum_l(KG_l^2/\Delta_l)\sin\Delta_l\tau$. At the next order the friction $-Kr^{(1)}w_j^{(1)}$ drives every mode. Its
constant part, which comes from $l=j$ alone, is $-K^3G_j^3/2\Delta_j^2$, and it shifts the average of the mode by this
amount divided by $\ii\Delta_j$. The parts at other frequencies average out, and the one at the frequency of the mode only
shifts that frequency. Hence, to relative order $K^2$,
\begin{equation}
\langle w_j\rangle=\frac{KG_j}{\ii\Delta_j}\Big(1-\frac{a_j}2\Big),\qquad \sum_j|\langle w_j\rangle|^2=\frac{K^2}{K_c^2}\Big(1-\frac{\sum_ja_j^2}{\sum_ja_j}\Big),
\label{SM-eq:orbit}
\end{equation}
with $a_j=K^2G_j^2/\Delta_j^2$. For the random context of \Mfig{fig:theory} this agrees with the integrated orbits within a
few per cent up to $K=0.85K_c$. An orbit thus keeps most of the weight of its centre, and the layer reads the centre,
Eq.~\eqref{SM-eq:silent}.

\subsection{Onset and one effective token}\label{SM-sec:onset}

Near the threshold only a few moments of the context matter, and the context acts as a single token. One token with the same moments $\sum_jG_j^2/\Delta_j^2$ and $\sum_jG_j^2/\Delta_j^4$ as the context has
\begin{equation}
\bar G=\frac{\sum_jG_j^2/\Delta_j^2}{\big(\sum_jG_j^2/\Delta_j^4\big)^{1/2}},\qquad
\bar\Delta=\Big(\frac{\sum_jG_j^2/\Delta_j^2}{\sum_jG_j^2/\Delta_j^4}\Big)^{1/2},\qquad K_c=\frac{\bar\Delta}{\bar G}.
\end{equation}
Expanding the inverse of the left side of Eq.~\eqref{SM-eq:norm} to first order in $(Kr)^2$ gives Adler's law of this token,
\begin{equation}
r=\bar G\sqrt{1-\frac{K_c^2}{K^2}}\qquad(K>K_c),
\label{SM-eq:sqrt}
\end{equation}
\Meq{eq:sqrt}. It is exact for one token and near the onset, where $r\simeq\bar G\sqrt{2(K-K_c)/K_c}$ rises with Kuramoto's
exponent. Further from the onset the receiver recruits tokens that the effective token does not see, and the root of
Eq.~\eqref{SM-eq:norm} lies above Eq.~\eqref{SM-eq:sqrt}, with $r\to(\sum_jG_j^2)^{1/2}>\bar G$ as $K\to\infty$. One Newton step on
$\ln\sum_ja_j=0$ in $\ln\mu$, $\mu=Kr$, recovers most of the difference,
\begin{equation}
\mu\to\mu\Big(\sum_ja_j\Big)^{1/(2\kappa)},\qquad \kappa=\frac{\sum_ja_j\,\mu^2/(\mu^2+\Delta_j^2)}{\sum_ja_j}.
\label{SM-eq:step}
\end{equation}

Contexts differ in how many tokens they recruit, and to compare them we collapse the order parameter. With $F(\mu)=\sum_jG_j^2/(\mu^2+\Delta_j^2)$, Eq.~\eqref{SM-eq:norm} reads $K^2F(Kr)=1$ and
Eq.~\eqref{SM-eq:kc} reads $F(0)=K_c^{-2}$. The order parameter
\begin{equation}
\hat r=\Big[1-\frac{F(Kr)}{F(0)}\Big]^{1/2}
\label{SM-eq:rhat}
\end{equation}
vanishes with $r$ and equals $\sqrt{1-K_c^2/K^2}$ at the root, for every context. It removes the recruitment, which differs
from context to context, and Fig.~\ref{fig:theory}(a) of the main text compares contexts through it.

\subsection{Linewidth}\label{SM-sec:linewidth}

At $K_c$ the derivatives of $r$ diverge, which hampers training. We therefore give every token a phase noise: in the frame
of the dial, its phase is $\Delta_j\tau+\xi_j(\tau)$, with $\xi_j$ a Brownian motion of diffusion constant $D$. At constant
friction the amplitude of a token is a memory of its past, weighted by the decay of the mode,
$w_j(\tau)=KG_j\int_0^\infty e^{-(Kr+\ii\Delta_j)s}\,e^{-\ii[\xi_j(\tau)-\xi_j(\tau-s)]}\,ds$, and the noise averages the phase
factor to $e^{-Ds}$. Every line becomes a Lorentzian of half-width $D$, and
$w_j=KG_j/(\mu+\ii\Delta_j)$ with $\mu=Kr+D$. The order parameter, $r=\sum_jG_j\Re w_j$, then gives $Kr=\mu\sum_ja_j$, that is,
\begin{equation}
\mu\Big[1-\sum_j\frac{K^2G_j^2}{\mu^2+\Delta_j^2}\Big]=D,\qquad \sum_ja_j=1-\frac D\mu .
\label{SM-eq:noise}
\end{equation}
The norm that the weights miss, $D/\mu$, is carried by the incoherent part of the receiver that the noise creates.
The condition has exactly one root $\mu>D$ at every coupling, a smooth function of all parameters. The transition is rounded, and as
$D\to0$ the two branches come back. The layer uses $D=10^{-3}K$. The root is found by bisection on
$(D,D+K(\sum_jG_j^2)^{1/2}]$, and one Newton step at the root gives its exact derivatives.

Without a loop, the effective token with a linewidth solves the cubic
\begin{equation}
(\mu-D)(\mu^2+\bar\Delta^2)=K^2\bar G^2\mu ,
\label{SM-eq:cubic}
\end{equation}
which has exactly one root above $D$ (Cardano). It is refined by the Newton step of Eq.~\eqref{SM-eq:step} with the linewidth,
$\mu\to\mu\exp\{[\ln\sum_ja_j-\ln(1-D/\mu)]/[2\kappa+D/(\mu-D)]\}$.

\subsection{Relation to Kuramoto's model}\label{SM-sec:kuramoto}

The radio head borrows Kuramoto's language, but the setting differs in two ways worth making explicit. In Kuramoto's model
$N$ phases are coupled all to all, and the threshold exists for $N\to\infty$~\cite{S-kuramoto,S-acebron}. Here the tokens do not respond. The coupling is a directed star, and the transition belongs to one receiver driven by a
fixed population. It exists at finite $N$, and the content of the context sets it. The onset has Kuramoto's exponent. The curve $\sqrt{1-K_c^2/K^2}$ and the
$\sqrt{1-K_c/K}$ of a Lorentzian distribution of frequencies share this exponent, and near the onset the first is larger by a factor $\sqrt2$.

For $m=1$ the receiver and the tokens are phases, $\bz=e^{\ii\phi}$ and $\bx_j=e^{\ii\theta_j}$, and Eq.~\eqref{SM-eq:lohe} becomes
\begin{equation}
\dot\phi=-\alpha+K\sum_jG_j\sin(\theta_j-\phi),\qquad \dot\theta_j=-\varepsilon\,(i-j),
\label{SM-eq:phase}
\end{equation}
Kuramoto's equation for one oscillator, whose natural frequency $-\alpha$ is the dial, driven by a fixed population whose
frequencies are the ages. Token $j$ is detuned from the receiver by $\Delta_j=\alpha-\varepsilon(i-j)$, as in the full model. This is the model one would write with phases alone,
and where it fails shows what the dimension is for. A phase carries no content, since $|\langle\bq|\bv_j\rangle|=1$ for every
token, so the antenna passes all of them and the gains would have to be put in by hand. And even with the gains given, a
phase follows one frequency at a time. For a single token, the phase difference $\psi=\theta-\phi$ obeys Adler's
equation~\cite{S-adler}, $\dot\psi=\Delta-KG\sin\psi$. It locks inside the range $|\Delta|\le KG$, with $\sin\psi=\Delta/KG$, and
outside it drifts, with an in-phase part that averages to zero and a quadrature part~\cite{S-thesis}
\begin{equation}
\langle\sin\psi\rangle=\frac{\Delta-{\rm sgn}(\Delta)\sqrt{\Delta^2-K^2G^2}}{KG}\simeq\frac{KG}{2\Delta}.
\label{SM-eq:adler}
\end{equation}
Far from resonance this is, up to a factor of two, the quadrature tail $KG/\Delta$ of the Lorentzian of Eq.~\eqref{SM-eq:w}.
Close to it the two differ. With several tokens in range a phase locks to one of them, chosen by where it starts, and
nothing is left of superposition, normalization or silence. In $\mathbb C^m$ tokens with nearly orthogonal contents take
nearly orthogonal directions of the receiver, which can follow all their frequencies at once, and the incoherent sum of
their gains sets a single threshold for the context.

\subsection{Predictions and capacity}\label{SM-sec:props}\label{SM-sec:capacity}

\emph{Normalization} is a result: $\sum_ja_j=1-D/\mu$, which tends to one above threshold and to $K^2/K_c^2$ below.
\emph{Abstention.} A query whose antenna finds no matching content has small gains, a large $K_c$ and a total weight
$(K/K_c)^2\ll1$. Its output is then small, rather than an average of the context. The missing norm stays in the memory $\bc$ of the receiver,
which is not read, so no sink token is needed, and the residual stream carries the token on. \emph{Copies.} Below threshold, Eq.~\eqref{SM-eq:silent} is a sum over
tokens, so $k$ copies of gain $G$ add one term each, until they bring the receiver to threshold and the total weight
saturates at one. Softmax gives the copies the fraction $ke^s/(ke^s+Z_0)$ (Sec.~\ref{SM-sec:others}), which depends on the length unless a sink dominates $Z_0$. \emph{Extrapolation.}
Position enters through the detunings, a rule of the age valid at any length. The Lorentzian tail falls as the inverse square of the age, so $\sum_j\Delta_j^{-2}$ converges and old tokens
add a bounded background. A token is read only inside the locking range, $|\Delta|<KG$, so a receiver reaches back to the
age $(\alpha+KG)/\varepsilon$, beyond which the weight of a token falls as $(KG/\Delta)^2$.

\emph{Capacity.} A random content has $|\langle\bq|\bv\rangle|^2\simeq1/m$. With $n=1$, about $m$ unrelated tokens in one band
carry as much gain as one matching token, so a head resolves about $m$ tokens at once. The same overlap sets the error of
the closed form: wide heads make the layer both exact and selective. A directive antenna lowers the gain of a random
content to $m^{-n}$. A head of ageing rate $\varepsilon$ holds about $Kr/\varepsilon$ tokens in its band, which calls for
$n\gtrsim\ln(Kr/\varepsilon)/\ln m$. Trained language models are consistent with this [inset of Fig.~\ref{fig:physics}(d) of the main text]: $n$ grows linearly
with $\ln(1/\varepsilon)$ across the heads of two seeds (correlation 0.97), from 0.4--0.8 in the fast heads of the first layer
to 3.5--3.9 in the slowest head of each seed ($\varepsilon\simeq0.01$). The slow heads are also the most strongly locked, and they read by
content: deep in the locked phase $Kr\gg|\Delta_j|$ over most of the context, and $w_j\simeq G_j/r$ no longer depends on the age.

\section{Closed form against the dynamics}\label{SM-sec:check}

The closed form rests on one approximation, the averaging of the beats. Here we test it directly: we integrate
Eq.~\eqref{SM-eq:frame} without noise and compare the time-averaged amplitudes with the closed form (\Mfig{fig:theory}).

\emph{Integration.} Fourth-order Runge--Kutta, with the receiver renormalized at every step, time in units of $1/K_c$, a step
$\min(0.01,0.12/\max_j|\Delta_j|)$ and averages over the second half of the run. Runs last at least 300 time units and long
enough for the averaging half to hold ten periods of the slowest token. Every receiver is integrated at twelve couplings from
$K=0.5K_c$ to $2.5K_c$, from two (trained contexts) or three (random contexts) random starts. Tokens with $|\Delta_j|>20K_c$,
whose weight is below $1/400$ of a resonant one, are left out.

\emph{Random contexts} (insets). Sixteen tokens with a fixed profile of gains and detunings and random contents in $\mathbb C^m$,
$m=4$ to 1024, ten draws of the contents per width. Above threshold the order parameter follows the root of Eq.~\eqref{SM-eq:norm} and the weights fall on the
Lorentzian. Their error falls as $1/m$ over the whole range [inset of Fig.~\ref{fig:theory}(c) of the main text]. Below threshold the receiver
orbits the silent state, with the total weight of Eq.~\eqref{SM-eq:orbit}.

\emph{Trained contexts} (main panels). We take the gains, detunings and contents of receivers of the second layer of the
language models of Fig.~\ref{fig:theory}(d,e) of the main text, at random positions of the validation stream: twenty resolved receivers per width,
and fourteen drawn without condition. Trained contents are correlated, unlike random ones.

\emph{Resolved and unresolved receivers.} The averaging of Sec.~\ref{SM-sec:solution} needs the beats between tokens, at
multiples of $\varepsilon$, to be faster than the band of the receiver, $Kr$. A slow head can hold tens of tokens inside one
band. It then hears them as one field and cannot tell them apart in frequency. Over the sweep, $K\le2.5K_c$, the band $Kr$ stays within a few $K_c$, so we call a receiver resolved when
$\varepsilon\ge K_c$, a condition fixed before the comparison, and show both kinds in Fig.~\ref{fig:theory}(c) of the main text. The error of the
resolved receivers falls with the width as $Kr/(m\varepsilon)$: a beat at the
spacing $\varepsilon$ passes the band with a relative amplitude $Kr/\varepsilon$ and couples two tokens through an overlap whose
square is of order $1/m$. Their order parameter $\hat r$ follows the root, $\sqrt{1-K_c^2/K^2}$, closely once the receiver is clearly locked. Unresolved
receivers have larger errors, and some lock below $K_c$: their tokens act as one coherent field, stronger than the
incoherent sum that sets $K_c$.

In the language model of Fig.~\ref{fig:physics}(d) of the main text a large part of the receivers of the first layer are resolved, at most a
quarter of those of the second layer, and none of the slowest head. Integrating receivers at the trained coupling, without
condition, the output of a head differs from the closed form by 10--20\,\%, with no growth in the number of tokens in the
band: the beats move weight between neighbours of nearly equal weight and leave their sum in place. The closed form is the
stationary state of resolved receivers and, within this error, the mean field of the others. A physical receiver would need
a reading time of order $1/\varepsilon$ to resolve a slow head.

\section{Softmax and the layer of Pasqualetti and Guo}\label{SM-sec:others}

\emph{Softmax with rotary position.} Rotary position~\cite{S-rope} rotates query and key plane by plane,
$\bq_i\to R_i\bq_i$, $\bk_j\to R_j\bk_j$, $R_t=\mathrm{diag}(e^{\ii\beta_\nu t})$, $\beta_\nu=10^{-8\nu/d}$, so that the score
$s_{ij}=\Re\langle R_i\bq_i|R_j\bk_j\rangle/\sqrt d$ depends on the lag only, a sum of oscillators of fixed frequencies $\beta_\nu$,
one per plane. The weights $a_{ij}=e^{s_{ij}}/\sum_le^{s_{il}}$ are the Gibbs distribution of the energies $-s_{ij}$; there is
no dynamics. They sum to one for every query, so a query that matches nothing returns an average. On $k$ copies of score $s$,
with the rest of the context of partition function $Z_0$, the weight is
\begin{equation}
A_k=\frac{ke^{s}}{ke^{s}+Z_0},
\label{SM-eq:smcopies}
\end{equation}
a fraction of the context rather than a count. The rotations are fixed functions of the position, and lags beyond the trained length produce
scores the network has not seen.

\emph{Pasqualetti and Guo (PG).} Every position $j$ owns a learned anchor $\bmm_j$ on the unit sphere, the query couples to
it with $W_{ij}={\rm softplus}(\langle\bq_i,\bk_j\rangle/\sqrt d)$, and an oscillator $\bm y_i$ without natural frequency
relaxes in the static field $\bh_i=\sum_jW_{ij}\bmm_j$~\cite{S-pg},
$\dot{\bm y}_i=\bh_i-\langle\bh_i,\bm y_i\rangle\bm y_i$, a gradient flow with the fixed point $\bm y_i^*=\bh_i/|\bh_i|$. The layer
reads it against the anchors, $a_{ij}\propto1+\langle\bm y_i^*,\bmm_j\rangle$, normalized. Content is mapped to position through
the anchors and back to weights through the readout; the coupling is computed digitally and the oscillator only turns it
into a direction. The weights sum to one, a query that matches nothing returns an average, copies enter only through the
direction of $\bm y_i^*$, and the anchors are learned for the trained positions and do not extrapolate.

The radio head grew out of the bachelor thesis of one of us~\cite{S-thesis}, which gave the oscillator of Pasqualetti and Guo
a natural frequency. That work kept the Lohe dynamics on the sphere, moved to the frame in which the anchors are static, and
reduced the high-dimensional equation exactly to Adler's equation in the plane of the pull. Its closed form already held
two pieces of the radio head. Locked, the probe settles with a phase lag that moves attention away from the exact match and
acts as a relative positional code. Drifting, its time average shrinks as $KG/2|\Delta|$ and points in quadrature with the
pull [Eq.~\eqref{SM-eq:adler}], so a probe that cannot lock abstains. The thesis also read the state through a Lorentzian, the
response of a physical resonance, and saw the gap to softmax close as heads widen. The radio head moves the state from the positions to the contents and the position
into the frequencies of the tokens. Without ageing ($\varepsilon=0$) its field would be static,
$\bh(\tau)=e^{\ii\alpha\tau}\sum_jG_j\bv_j$, and Eq.~\eqref{SM-eq:frame} would reduce exactly to Adler's equation with the coherent
gain $|\sum_jG_j\bv_j|$: a single field, which cannot separate tokens. Distinct detunings replace the coherent gain by the
incoherent one, $(\sum_jG_j^2)^{1/2}$. It is the ageing that lets the receiver tell tokens apart.

\emph{Other layers.} Each ingredient of softmax has its patches, and studying them is not the aim of this work. As a check, we trained a few of them with the same protocol, on the synthetic tasks and, for linear biases, on language. Linear biases~\cite{S-press}, which subtract from each score a slope times the age, extrapolate as well as the radio head or better once the slope is learned: the first layer learns a steep slope, a clock on the previous token, and the second an almost flat one that reads by content, the same division of labour that the radio head finds. With the published slopes they also extrapolate in language. Layers without normalization, such as sigmoid attention~\cite{S-ramapuram}, softmax with one added to its denominator~\cite{S-miller} or gated softmax~\cite{S-qiu}, can likewise return little when nothing matches and let copies add up; sigmoid attention, for instance, keeps its count at longer lengths, although it learns neither recall nor induction. What these layers do not give is a reason for the decay or for the silence, nor a threshold, set by the context, that predicts where the network fails.

\begin{table}[ht]
\centering
\small
\begin{tabular}{llll}
\toprule
 & softmax + RoPE & Pasqualetti--Guo & radio head \\
\midrule
state & --- & position & content \\
position & rotation of $\bq$, $\bk$ & anchors & frequency (age) \\
dynamics & none & gradient flow & partial locking \\
weights & Gibbs & fixed-point readout & $KG_j/(Kr+\ii\Delta_j)$ \\
normalization & $\sum a=1$ & $\sum a=1$ & $\sum|w|^2=1$ locked, $<1$ silent \\
no match & average & average & silence \\
$k$ copies & $ke^s/(ke^s+Z_0)$ & bounded readout & linear in $k$ below $K_c$ \\
\bottomrule
\end{tabular}
\caption{Three attention layers.}
\label{SM-tab:comparison}
\end{table}

\section{Implementation and methods}\label{SM-sec:impl}\label{SM-sec:methods}

This section describes the layer as it is coded and how every network was trained. Table~\ref{SM-tab:settings} gathers the
settings.

\emph{The layer.} Per head of width $d=2m$: contents and antennas $\bv_j=W_V\be_j/|W_V\be_j|$ and $\bq_i=W_Q\be_i/|W_Q\be_i|$;
gains $G_{ij}^2=\rho+(1-\rho)|\langle\bq_i|\bv_j\rangle|^{2n}$, with a small learned floor $\rho$, which lowers the threshold of a nearly resonant token, so that abstaining needs $K^2\rho$ below the squared detuning of the nearest token, at most $(\varepsilon/2)^2$; detunings
$\Delta_{ij}=\alpha_i-\varepsilon(i-j)$ with $\alpha_i=\bm w_\alpha^\top\be_i+b_\alpha$. The width $\mu_i$ solves Eq.~\eqref{SM-eq:noise}, by
20 bisections without gradients and one Newton step with them. The implicit function theorem then gives the exact
derivatives, and training is ordinary backpropagation through the closed form. The weights are
$w_{ij}=KG_{ij}/(\mu_i+\ii\Delta_{ij})$ and the output is Eq.~\eqref{SM-eq:out}. The one-token version takes $\mu_i$ from
Eq.~\eqref{SM-eq:cubic} instead, with $\bar\Delta_i^2=M_i/N_i$, $K^2\bar G_i^2=M_i^2/N_i$, $M_i=\sum_jK^2G_{ij}^2/\Delta_{ij}^2$ and
$N_i=\sum_jK^2G_{ij}^2/\Delta_{ij}^4$, and frequencies rescaled to keep the cubic finite in single precision. The learned
parameters are $W_Q$, $W_V$, $\bm w_\alpha$, $b_\alpha$, $W_O$ and, per head, $n$, $\rho$, $\varepsilon$, $K$ and $g$. Each head has its
own receivers. The outputs of the heads are concatenated and mapped by $W_O$, as in any transformer. The cost is $O(T^2d)$ per
head, as for softmax, with a constant factor from the bisection and the complex arithmetic: a run took about 1.7 times as long as softmax in language and three times in the synthetic tasks.

\emph{Initialization.} Every head starts at $\varepsilon=T^{-1/2}$, the same in all heads of a layer, so that their spread
after training is learned. It listens at the age 1.5 ($\alpha=1.5\varepsilon$), with $n=1$, $g=1$ and $\bm w_\alpha=0$, and starts
locked at $K=1.5K_c$ for a context of random contents, $K=1.5\,\varepsilon\,m^{1/2}[\sum_{L=1}^{T-1}(L-1.5)^{-2}]^{-1/2}$. The
counting layer starts at $K=K_c$, since the task needs the silent phase.

\emph{Host network and protocol.} Pre-norm transformer blocks with a feed-forward layer and residual connections. The
embedding, optimizer, schedule and data are the same for every attention layer; only the layer changes. AdamW, weight decay
$10^{-4}$ on the matrices, gradient clipping at 1, linear warm-up over 2\,\% of the steps, constant rate, and linear decay to
zero over the last 20\,\%. Nothing is chosen on the test set. The oscillator of PG has the width of the head and the
sinusoidal input code of the published layer, and our implementation was checked against their code.

\emph{Synthetic tasks.} Two layers with one head, $d_{\rm model}=64$, a feed-forward layer of width $2d_{\rm model}$ and no
dropout. Every epoch draws new sequences. Training runs for 80 epochs and keeps the epoch of best validation accuracy, with
five seeds. The learning rate is the best validation among $3\times10^{-4}$, $10^{-3}$ and $3\times10^{-3}$ at seed 0, which is
$10^{-3}$ for every layer except two that do not learn at it, softmax in induction and PG in recall, which use
$3\times10^{-4}$.

\emph{Language.} All of TinyStories~\cite{S-eldan}, one pass, with the vocabulary and split of Ref.~\cite{S-pg}. Two layers, a
feed-forward layer of width $2d_{\rm model}$, dropout 0.1, trained length 128 and learning rate $10^{-3}$, which beats
$5\times10^{-4}$ for every layer after one epoch (Sec.~\ref{SM-sec:results}). Three families of networks: $d_{\rm model}=128$
split into one to eight heads [Fig.~\ref{fig:theory}(d) of the main text], four heads of width $2m$ with $d_{\rm model}=8m$
[Fig.~\ref{fig:theory}(e) of the main text], and $d_{\rm model}=256$ with four heads (\Mfig{fig:physics}). Two seeds each, one at $m=128$ in the four-head series.

\emph{Tasks.} Recall~\cite{S-arora}: $P$ key--value pairs followed by 16 queries, 20\,\% of them for unlisted keys, length
$T=2P+16$, trained at $P=16$ and tested up to $P=496$ (length 1008). Counting: the number (0--15) of earlier copies of the query
token, read as a number (one output, squared error, right within 0.5), trained at length 32 and tested up to 512. Induction~\cite{S-olsson}:
the token that followed the previous occurrence of the query, trained at length 32 and tested up to 1024.

\begin{table}[ht]
\centering
\small
\begin{tabular}{lll}
\toprule
 & synthetic tasks & language (TinyStories) \\
\midrule
layers, heads & 2, 1 & 2, 1 to 8 \\
$d_{\rm model}$, feed-forward width & 64, 128 & 64 to 1024, $2d_{\rm model}$ \\
dropout & 0 & 0.1 \\
vocabulary & 51 (recall: 577) & 8004~\cite{S-pg} \\
training length & 32 (recall: 48) & 128 \\
training data & 20\,000 new sequences per epoch & $4.6\times10^6$ sequences of 128 tokens, one pass \\
validation, test & 2000, 2000 sequences & split of Ref.~\cite{S-pg} \\
batch, steps & 128, about $1.25\times10^4$ (80 epochs) & 256, $1.8\times10^4$ \\
parameters & $6\times10^4$ to $10^5$ & $1.1\times10^6$ to $3.3\times10^7$ \\
hardware, time per run & CPU (two threads), 15--80 min & Apple M4 Pro GPU, 1--13 h \\
\bottomrule
\end{tabular}
\caption{Settings of the trained networks. The optimizer and schedule are the same in both (see text).}
\label{SM-tab:settings}
\end{table}

\emph{Reading the receivers.} For every query of a trained radio head we compute $G_{ij}$, $\Delta_{ij}$, $K_c$ and $r$
from the learned parameters and the inputs of the layer, which tells in which phase each receiver works. Section~\ref{SM-sec:check}
integrates Eq.~\eqref{SM-eq:frame} with the same $G_{ij}$, $\Delta_{ij}$ and $\bv_j$.

\section{Checks and controls in the trained networks}\label{SM-sec:results}

\subsection{Each task against the phase diagram}\label{SM-sec:taskchecks}

For each task we read the receivers of the trained network as in Sec.~\ref{SM-sec:impl} and compare them with the phase
diagram. The figures of the letter show the main comparisons; here we add what they leave out.

\emph{Recall.} The order parameter of the retrieval layer, in units of its strong-coupling limit $r_\infty=(\sum_jG_j^2)^{1/2}$, follows Eq.~\eqref{SM-eq:sqrt}
for nearly every query, since a single key dominates the field and $\bar G\simeq r_\infty$ [Fig.~\ref{fig:physics}(a) of the main text]. With four and eight times the trained number of pairs the radio head keeps an
accuracy of 0.94 and 0.74, against 0.79 and 0.39 for softmax, while PG drops to chance. The errors of the radio head come from many detuned keys
that together push an unlisted key over the threshold.

\emph{Counting.} All layers learn the task, PG less well (test accuracy 0.87--0.98). The total weight of the counting layer
equals $(K/K_c)^2$ query by query and grows as $k$ times the weight of one copy, also beyond the 15 copies of training:
training keeps the receiver below threshold. Raising the coupling of a trained network, $K\to cK$ with $g\to g/c$, leaves the layer unchanged
below threshold and brings the threshold down to fewer copies. The count is then lost exactly there: over factors $c$ from 1 to 40, the network
counts right nearly always below threshold and almost never above. In longer contexts the weight on the copies follows the prediction from
the weight of one copy against its age, while that of softmax falls as $32/T$.

\emph{Induction.} The first layer becomes a clock that listens to the previous token, with its dial at an age close to
one. The second layer listens further back, and its weight on the answer follows $\min[1,(KG/\Delta)^2]$ on both sides of
the edge of the locking range [Fig.~\ref{fig:physics}(c) of the main text]. Beyond the trained length the accuracy equals the fraction of answers
inside the locking range in every seed [Fig.~\ref{fig:physics}(g) of the main text]. The spread between seeds is physical. A receiver reaches back
to the age $(\alpha+KG)/\varepsilon$. Four seeds learn a reach of about 50 tokens. The fifth learns a stronger coupling, a
sharper antenna and a slower dial, reaches about 170 tokens, and keeps nearly all its accuracy at four times the trained length,
where the others lose about half. Training at $T=32$ cannot tell the two solutions apart.

\emph{Language.} All heads start at $\varepsilon=T^{-1/2}\simeq0.09$, and training spreads them from 0.01 to 0.55. The heads
split into three regimes. The slowest head locks deepest (median $K/K_c$ of 10--20), has the sharpest antenna and reads by
content. The fast heads read by position, closer to threshold (median $K/K_c$ of 2--5). In one seed a head keeps over a third of
its receivers silent. On contexts cut from the concatenated validation stories (mean of two seeds; for PG, positions beyond 128 get fresh
anchors from the initialization):

\begin{center}
\small
\begin{tabular}{lccccc}
\toprule
context length & 64 & 128 (trained) & 256 & 512 & 1024\\
\midrule
perplexity, softmax + RoPE & 7.71 & 6.93 & 15.6 & 46.4 & 79.9\\
\phantom{perplexity,} radio head & 7.64 & 6.99 & 7.00 & 7.65 & 9.42\\
\phantom{perplexity,} PG & 8.91 & 8.56 & 38.0 & 79.2 & 123\\
accuracy, softmax + RoPE & 0.532 & 0.550 & 0.427 & 0.285 & 0.209\\
\phantom{accuracy,} radio head & 0.531 & 0.545 & 0.542 & 0.529 & 0.501\\
\phantom{accuracy,} PG & 0.502 & 0.508 & 0.330 & 0.238 & 0.187\\
\bottomrule
\end{tabular}
\end{center}

\noindent These contexts concatenate unrelated stories, so a layer whose weights decay with the age, by hand or by resonance, needs little beyond the current story. The radio head keeps its perplexity up to twice the trained length because $\sum_jG_{ij}^2/\Delta_{ij}^2$ converges as
the tokens age, with the inverse square of the age: the receivers barely change when old tokens are added.

\subsection{Width, and the one-token approximation}\label{SM-sec:width}

\noindent Next-token accuracy relative to softmax, test split, mean of two seeds (one at $m=128$ with four heads). The one-token version replaces the exact root by the cubic of
Sec.~\ref{SM-sec:onset} and leaves the rest of the layer as it is:

\begin{center}
\small
\begin{tabular}{llccccc}
\toprule
 & & $m=8$ & 16 & 32 & 64 & 128\\
\midrule
$d_{\rm model}=128$ [Fig.~\ref{fig:theory}(d) of the main text] & radio head & 0.978 & 0.984 & 0.990 & 1.006 & \\
 & one effective token & 0.974 & 0.982 & 0.989 & 1.005 & \\
 & PG & 0.961 & 0.965 & 0.969 & 0.969 & \\
four heads [Fig.~\ref{fig:theory}(e) of the main text] & radio head & 0.971 & 0.984 & 0.991 & 0.993 & 0.997\\
 & one effective token & 0.967 & 0.982 & 0.987 & 0.989 & 0.984\\
 & PG & 0.941 & 0.965 & 0.973 & 0.975 & 0.977\\
\bottomrule
\end{tabular}
\end{center}

\noindent In the four-head series, where $d_{\rm model}=8m$, the test loss of softmax falls from 2.50 to 1.62 and the excess of
the radio head from 0.080 to 0.009 nats. At $m=128$ the effective token falls behind the root. Its width matches the root near threshold but is clearly off for
the receivers deep in the locked phase ($K/K_c>5$) of one head of this model, where the one-token approximation is not meant
to hold.

\subsection{Controls of the protocol}\label{SM-sec:controls}

\noindent We checked that the comparison does not hinge on our protocol. At $d_{\rm model}=128$ with four heads, we trained
every layer at a lower learning rate and with the protocol of Ref.~\cite{S-pg}, five epochs at $5\times10^{-4}$:

\begin{center}
\small
\begin{tabular}{lccc}
\toprule
perplexity & softmax + RoPE & radio head & PG \\
\midrule
one epoch, $10^{-3}$ (test) & 9.04 & 9.55 & 10.02 \\
one epoch, $5\times10^{-4}$ (test) & 10.00 & 10.50 & 11.99 \\
protocol of Ref.~\cite{S-pg} (validation) & 8.81 & 9.25 & 9.54 \\
reported in Ref.~\cite{S-pg} (validation) & 8.67 & --- & 9.11 \\
\bottomrule
\end{tabular}
\end{center}

\noindent In every protocol the radio head stays about 0.05 nats from softmax, and PG further behind.

\end{document}